\documentclass[a4paper,10pt]{ijuc}

\usepackage{fancyhdr}

\usepackage{graphicx} \usepackage{subfigure} \usepackage{stmaryrd}

\usepackage{booktabs}

\usepackage{amsfonts} \usepackage{amssymb}

\usepackage[colorlinks=true, pdftex=true, pdfstartview=FitV,
linkcolor=black, citecolor=black, urlcolor=blue]{hyperref}

\newcommand{\superscript}[1]{\small\ensuremath{^\textrm{#1}}\normalsize\hspace{0.25ex}}

\renewcommand{\th}[0]{\superscript{th}}

\newcommand{\lips}{\ldots}

\begin{document}


\title{From Hypocomputation to Hypercomputation} 
\author{David Love\email{d.love@shu.ac.uk}} 
\institute{School of Computing and Management Sciences,\\
Sheffield Hallam University, City Campus,\\ Howard
Street, Sheffield, S1 1WB.}

\markboth{Special Edition on Hypocomputation} 
{David Love --- From Hypocomputation to Hypercomputation}

\def\received{Received \today; }

\maketitle

\begin{abstract}

Hypercomputational formal theories will, clearly, be both structurally
and foundationally different from the formal theories underpinning
computational theories. However, many of the maps that might guide us
into this strange realm have been lost. So little work has been done
recently in the area of metamathematics, and so many of the previous
results have been folded into other theories, that we are in danger of
loosing an appreciation of the broader structure of formal theories.

As an aid to those looking to develop hypercomputational theories, we
will briefly survey the known landmarks  both inside and outside the
borders of computational theory. We will not focus in this paper on
\emph{why\/} the structure of formal theory looks the way it does.
Instead we will focus on \emph{what\/} this structure looks like, moving
from hypocomputational, through traditional computational theories, and
then beyond to hypercomputational theories.

\end{abstract}


\section{Introduction}\label{s:Introduction}

\subsection{Returning to Metamathematics}\label{ss:RetToMet}

Our main guide to the broader structure of formal theory comes from
research in the field of metamathematics. Very loosely, the field of
metamathematics tries to ``apply mathematics to itself'': using
only rigorous mathematical theories to probe the structure and foundations of
mathematics. Or at least it did once.

Throughout the late nineteenth and early twentieth centuries,
metamathematics was a very active field. The search for the foundations
of mathematics was on, and those mathematicians and logicians involved
in research in metamathematics were leading the charge.

Ultimately, that particular search for a foundational theory of
mathematics failed. Nonetheless, research in metamathematics has
provided the foundation for many modern theories of mathematics ---
particularly the formal theories of computation.

\medskip

We are not, though, aiming to
present a metamathematical map of formal theories in all their technical
glory here. Rather our aim is to highlight the main results, hoping to
make them accessible to the general body of researchers engaged in
hypercomputational research. In essence, we aim to point out useful
structures and foundations which may not be obvious at first sight to
those only familiar with computational theories. By exposing the broader
structures of formal theory, we hope to prompt further discussion of
hypercomputation --- and possibly even a renewal of interest in
metamathematics.

\medskip

Undeniably, metamathematics is not now a particularly active field.
Nonetheless, most of our understanding of formal theory (especially the
broader structures) comes from this field. Since the Second World War,
however, many metamathematical results have been re-cast as
computational theories. But in the translation, some important details
have been lost.

The recent focus on computational theory leaves us with two problems
when we return to a more general discussion of formal theories. Firstly,
many of the best known metamathematical results are known only through
the lens of computational theory. Secondly, much of the terminology in
computational theory is derived from the metamathematical theory.

Individually these difficulties are minor. Together, though, they can
obscure important details in the structure of formal theories.

\medskip

For instance, many presentations of computational theory make use of
abstract machines first proposed by Alan Turing in his 1936 paper
\emph{On Computable Numbers, With An Application To The
\emph{Entscheidungsproblem\/}\/}~\cite{Gandy:Confluence,
Turing:Computable}. We will not go into the details of the machines
proposed by Turing here\footnote{Readers looking for more details of
these machines may want to briefly review \S\ref{s:Computation} first.},
other than to note they are universally popular as means of presenting
results in computational theory. Although universal, the machines known
as `Turing Machines' in computational theory follow the pattern of one
of two machines types identified by Turing. In his original paper Turing
notes that~\cite[p. 232]{Turing:Computable}:

\begin{quotation}

For some purposes we might use machines (choice machines or
\emph{c\/}-machines) whose motion is only partially determined by the
configuration\lips. When such a machine reaches one of these ambiguous
configurations, it cannot go on until some arbitrary choice has been
made by an external operator. \lips\ In this paper I deal only with
automatic machines, and will therefore often omit the prefix
\emph{a\/}-.

\end{quotation}

For computational theory the distinction between the \emph{a\/}-machines
and \emph{c\/}-machines is irrelevant --- like Turing, computational
theory only considers the actions of automatic machines. Hence if we
confine our attention to computational theory, we can indeed forget all
about choice machines.

We must not forget, though, that Turing is writing in the context of
\emph{metamathematical\/} theories: not computational ones.
Metamathematical theories, though, can (and indeed do) allow very
different foundations for theories describing the actions of
\emph{c\/}-machines and those restricted to \emph{a\/}-machines.
Confusion is therefore bound to arise if we forget the context and treat
all `Turing machines' as automatic machines in the sense of
computational theories.

For instance, some researchers in computational theory \emph{define\/}
automatic machines as `\emph{deterministic Turing machines\/}' and
choice machines as `\emph{non-deterministic Turing
machines\/}'~\cite{Atallah:Algorithms}. This, though, requires the
underlying \emph{structure\/} of the choice facing the ``\emph{external
operator\/}'' to be \emph{exactly\/} the same as the structure of the
choice facing the machine. If accept this assumption, then we can
effectively ignore the operator: \emph{both\/} \emph{a\/}-machines and
\emph{c\/}-machines are then governed by the same structural theories.
Thus we can derive a `machine' theory the adequately covers both `kinds'
of machine.

These structural assumptions may seem reasonable. Indeed they are so
common that very few modern discussions even mention the possible
problems posed by \emph{c\/}-machines. But can we really assume the
structural assumptions underpinning \emph{a\/}-machines are
\emph{always\/} \textbf{exactly} the same as the structural assumptions
underpinning \emph{c\/}-machines?

\medskip

In computational theory the answer appears to be `\emph{yes}'. We know of no
case where the structural assumptions used for \emph{a\/}-machines fail
to adequately describe \emph{c\/}-machines.

However, in computational theories this answer is essentially a `proof by
definition'. We have defined \emph{c\/}-machines in terms of the
structural theories underpinning \emph{a\/}-machines. Hence we should not
reasonably expect any differences.

But what happens if we step outside the structural theories underpinning
\emph{a\/}-machines? Can we uncover a new set of structural assumptions
that allows a broader (but still rigorous) definition of a
\emph{c\/}-machines? Even if we found such a theory, how could we sensibly
relate it to the computational theory underpinning \emph{a\/}-machines?

\medskip

In this paper we contest that we can indeed find rigorous mathematical
theories that allow a sensible, but broader, definition of
\emph{c\/}-machines\footnote{As an illustration of what
a broader choice machine may look like, and particularly the foundations
of such machines, the reader is referred to \S\ref{s:Hypocomputation}
and \S\ref{s:Computation}}. Moreover, we can relate the these to the
computational theories underpinning \emph{a\/}-machines; thereby showing
how the structural assumptions of the \emph{c\/}-machines relate to
those of the \emph{a\/}-machines.

We can create these broader theories because metamathematical theories
allow a broader range of structural assumptions than simple computational 
theory. Computational theory is, after all, only
\emph{one\/} example of a formal theory permitted by metamathematical
methods and under metamathematical assumptions. However, discussing
these broader theories requires at least a passing familiarity with
metamathematical theory --- even if we seem to be covering the same
ground as computational theory. We
\emph{cannot\/} simply use computational theory as a substitute for
metamathematics.

\medskip

Nonetheless, given the paucity of activity in metamathematics since the
1950s, it seems unreasonable to expect the reader to be as familiar with
metamathematics as they might be with computation.  Readers who are
familiar with Hilbert's metamathematical program (and the main results)
might want to skip to the discussion on the structure of formal theories
in \S\ref{s:StructECM}. For the rest of us we will first look briefly at
the motivation for metamathematics: particularly the relationship
between metamathematics and the formal theories underlying computational
theory. From this brief examination, we can return to the main
discussion, again focusing on the metamathematical highlights, but
without going into the details. We will, though, point out sources for
those interested in further exploration.

\subsection{The Place of Formal Theory}

\subsubsection{Why `Metamathematics'?}\label{sss:WhyMeta}

In modern mathematical use, the term `formal' is most often used in
computational theories or modern logics. However, the term first
appeared in the late 19\th\ century, during the search for a grand
unified theory of mathematics. Although this search ultimately ended
without finding a grand theory, mathematicians engaged in the search
have provided us with most of our current understanding of the structure
of mathematics (and formal theories). In addition, this search produced
the branch of mathematics known as \emph{metamathematics}, from which
the founding theories of computation later emerged. Although
metamathematics has not received much in the way of in-depth study in
the last 60 years, if we are serious about understanding the foundations
of computational theories, metamathematics remains the best place to
start.

\medskip

Before we move onto the structure and assumptions of formal theory,
though, we ought to define what we mean by `metamathematics'. A full
introduction to this subject (and its history) is beyond the scope of
this paper\footnote{For an accessible introduction to metamathematics,
together with the main results and theories, see Stephen Kleene's
\emph{Introduction to Metamathematics}~\cite{Kleene:Introduction}.
Kleene also gives an excellent summary of the history and motivation for
metamathematics in Chapter III of his \emph{Introduction}, \emph{A
Critique of Mathematical Reasoning}. Another good commentary on the
early contenders for a full theory of mathematics is given by Kleene in
\emph{Mathematical Logic\/}~\cite[Ch IV, \S 36]{Kleene:Mathematical},}.
For our purposes here, we will use a later description of the scope and
aims of metamathematics, given by Stephen Kleene~\cite[p.
64]{Kleene:Introduction}:

\begin{quotation}

Metamathematics must study the formal system as a system of symbols,
etc.\ which are considered wholly objectively. This means simply that
those symbols, etc.\ are themselves the ultimate objects, are not being
used to refer to something other than themselves. The metamathematician
looks at them, not through and beyond them; thus they are objects
without interpretation or meaning.

\end{quotation}

Perhaps the most controversial statement in this definition is Kleene's
use of the term `formal' to describe the object of metamathematical
study. From our familiarity with computational theory and formal logic,
we are used to the term `formal' as the manipulation of ``\emph{objects
without interpretation or meaning\/}''. This modern sense of the term
`formal', though, usually only refers to the application of formal
assumptions to \emph{axiomatic systems\/}\footnote{An \emph{axiomatic\/}
system uses a finite number of \emph{axioms\/} (or \emph{postulates\/})
as founding assumptions and conditions for the mathematical system. The
(logical) consequences of these axioms are then used to develop a full
theory of the system~\cite[\S8]{Kleene:Introduction}. In most modern
axiomatic theories, the derivation from the axioms is achieved by
applying only those rules permitted by formal
theory~\cite{Kleene:Mathematical}. Hence the usual modern relationship
between axiomatic and formal theories.}. Kleene is using the the term
`formal' in the original sense of David Hilbert: and Hilbert, in turn,
was interested in the most basic properties of a mathematical theory. As
Kleene notes, in Hilbert's view a `formal theory' is slightly more
general than an axiomatic theory~\cite[p. 60]{Kleene:Introduction}:

\begin{quotation}

Since we have abstracted entirely from the content or matter, leaving
only the form, we say the original theory has been
\emph{formalized}.\lips\@ We say be reference to the form alone which
combination of words are sentences, which sentences are axioms, and
which sentences follow as immediate consequences from others.

\end{quotation}

Although this distinction may seem subtle, the modern use of the term
`formal' essentially eliminated metamathematics as a branch of serious
academic study.

\medskip

The problem is that all the classic formal theories (including
computation) turned out to be axiomatic theories. Moreover, even by the
end of the 1930s\footnote{Alonzo Church, creator of the first modern
computational theory, defined his famous thesis relating effective
decidable predicates (formal logic) to effectively calculable functions
(number theory) in
1936~\cite{Church:Unsolveable},\cite[\S60]{Kleene:Introduction}. This
thesis is now known as Church's Thesis (or, occasionally, the
Church-Turing Thesis after applying Church's arguments to the model of
computation developed by Alan Turing), and its proof remains one of the
central problems of modern
mathematics~\cite{Carnap:Logicist,Curry:Remarks} and
computation~\cite{Copeland:Church}.} mathematicians seemed to have found
a basic equivalence between (Hilbert's) formal theories and axiomatic
theories. Even more convincingly, all known equivalences relied on basic
properties in number theory; which many mathematicians regard as one of
the best described and understood branches of mathematics.

This seeming weight of evidence in favour of equating formal theory and
axiomatic theory led Alfred Tarski to denounce any separation between
the (formal) methods of David Hilbert's and the rest of
mathematics~\cite{Sinaceur:Tarski}:

\begin{quotation}

In [Tarski's] view, metamathematics became similar to any mathematical
discipline. Not only its concepts and results can be mathematized, but
they actually can be integrated into mathematics.\lips Tarski destroyed
the borderline between metamathematics and mathematics.

\end{quotation}

Although the opinion of Tarski and others led to a near abandonment of
metamathematics, this view implies a very precise structure of
mathematics. If Tarski's view is correct, then it would also seem to
rule out hypercomputational theories, since these break certain accepted
properties of formal theories\footnote{As we will examine in
\S\ref{s:Hypocomputation}}. Nonetheless, it appears Tarski's view of the
structure of mathematics is incorrect: at least if we believe the
metamathematical results of the 1930s.

Therefore, if we are interested in hypercomputation, we need to know
whether hypercomputational theories fit into the accepted structure of
mathematics. This means turning again to the metamathematics of the
1930s, to review the foundations of our current understanding.

\subsubsection{The Structure of Mathematics}

Trying to work out a structure for the whole of mathematical theory
would seem to be a monumental undertaking. Indeed most of our current
understanding evolved during 80 years of sustained effort, from the mid
19\th\ century to the 1930s. Fortunately, though, since most of the
results occur at the edges of formal theory, we are not required to know
them. By looking at the main results, we can, therefore, quickly build
up an map of most of the structures we are interested in.

We will start our exploration at the end of the 19\th\ century, with the
work of the German mathematician David Hilbert. Although David Hilbert
was not the first to try to find `the' founding theory of mathematics,
his work guided and shaped most of the programme from the late 19\th\
century to the middle of the 1930s. For Hilbert, a full guide to the
structure of mathematics ought to be divided into three distinct
``theories'', described by Kleene as follows~\cite[p.
65]{Kleene:Introduction}:

\begin{quotation}

In the full picture [of mathematics], there will be three separate and
distinct ``theories'': (a) the informal theory of which the formal
system constitutes a formalization, (b) the formal system or object
theory, and (c) the metatheory, in which the formal system is described
and studied.

\end{quotation}

\begin{figure}
   \centering \includegraphics{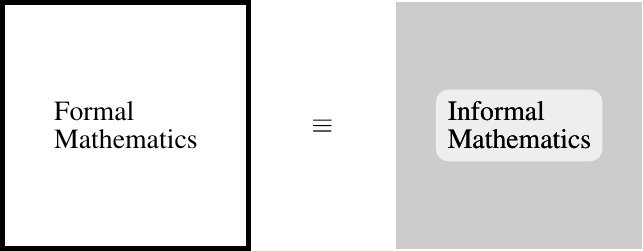}
   \caption{\label{figure:KMathStruct}A Na\"{\i}ve View of the Structure
   of Mathematics}
\end{figure}

Following this line of reasoning, and simplifying somewhat, we can draw
up a view of the structure of mathematics along the lines of
Figure~\ref{figure:KMathStruct}. Formal theory in
Figure~\ref{figure:KMathStruct} becomes a distinct region of
mathematics, with its own structure and identity. This does not mean
formal theory is the only possible source of mathematical theories,
though. Alongside formal theory we also have \emph{informal\/}
mathematical theories, again with its distinctive structures and a
separate identity. The challenge, then, is relating the formal and
informal theories of mathematics, without violating the structure and
identity of either. This is the task of the metatheory, shown as an
equivalence relation in Figure~\ref{figure:KMathStruct}. By carefully
constructing the metatheory, we hope to provide a foundation for the
formal theory --- without allowing the paradoxes and untidiness of the
informal to cross the divide\footnote{Unfortunately, though, it appears
that paradoxes are just as necessary to the foundations of formal theory
as they are to the informal. What we did learn from Hilbert's programme, 
though, was how to
create more limited formal theories, avoiding at least some of the
paradoxes. Of these limited formal theories, the axiomatic formal
theories have proved to be particularly useful.}.

\medskip

Metamathematics is, essentially, the creation and study of this
`equivalence' relation between formal and informal theory. Controversy
only arises when we ask ``\emph{what is the nature of this
metatheory\/}''?

Tarski argued strongly for the metatheory being a formal theory. If we
then place formal theory on an axiomatic foundation, we can go further
and argue for the elimination of the metatheory as a distinct theory.
For both the `metatheory' and the formal theory have the same
(or at least equivalent) foundation, and are governed by the same
(axiomatic) rules. Since one
theory can now be re-written in terms of the other, the distinction
`metatheory' and the formal theory seems (at best) academic.

Abandoning the distinction between the `metatheory' and formal theory
also allows us to abandon informal mathematical theories. If the old
`metatheory' relates informal and formal mathematics, by removing the
`metatheory' we can use (axiomatic) formal theory \emph{in place\/} of
informal theory. An (axiomatic) formal theory is always sufficient,
since every informal theory has a (axiomatic) formal equivalent.

\medskip

\begin{figure}
 \centering \includegraphics{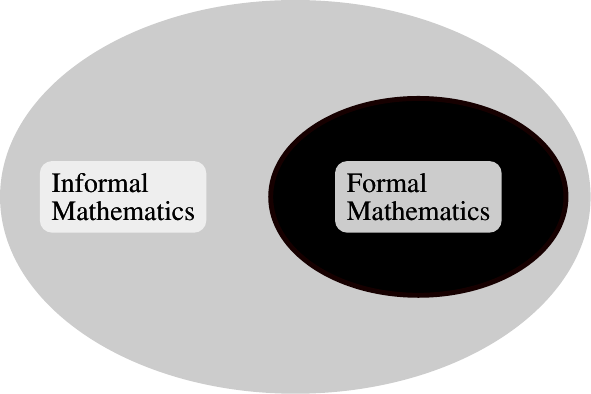}
 \caption{\label{figure:ThirtyStruct}The Structure of Mathematics at the
 End of the 1930s}
\end{figure}

Our problem, though, is that Figure~\ref{figure:KMathStruct} represents
an extreme simplification of the structure of mathematics as understood
at the end of the 1930s. It would be tempting to conclude from
Figure~\ref{figure:KMathStruct} that the region of formal mathematics
are the same `size' as the region of informal mathematics. If this were
the case, we might reasonably expect to find a formal equivalent of
every informal theory. However, by the end of the 1930s, the structure of
mathematics looked much more like Figure~\ref{figure:ThirtyStruct}.

Here, in Figure~\ref{figure:ThirtyStruct}, informal mathematics not only
covers the entire region of formal mathematics --- but quite a bit more.
In other words, Figure~\ref{figure:ThirtyStruct} suggests we might have
informal theories that cannot be fully formalised: some properties will
always remain beyond the borders of formal theory. Moreover, the
metatheory in Figure~\ref{figure:ThirtyStruct} \emph{cannot\/} simply be
eliminated. Instead the metatheory also has its own distinct identity
--- the metatheory becomes a `filter': condensing informal structures
into a shape acceptable to formal theory. In this case formal theory
might also be more general than simple axiomatic theories, since the
properties of the formal theory depend on the metatheory. We could
imagine situations where the metatheory allows an axiomatic statement of
a formal theory. In general, though, we are not \emph{limited\/} to
these situations.

For example, Georg Cantor's definition of a `set' attempts to capture in
a rigours theory our intuitive notions about `collections' of
`objects'~\cite[Chapter 1]{Kleene:Introduction}. Using Cantor's theory,
we seem to be able to capture and define mathematical notions, for
instance cardinality and the relationship between the natural and
rational numbers. It would appears, then, that Cantor's notion of a set
qualifies as a valid mathematical theory. Yet, when we attempt to define
certain mathematical notions using Cantor's theory, we run into a series
of paradoxes. Most famously, Bertrand Russel is identified with the
paradox that results when trying to define the boundary properties of
certain infinite sets permitted by Cantor's
theory~\cite[Russell's Paradox]{Clark:Paradoxes}.

In axiomatic set theories we attempt to avoid this paradoxes by
\emph{limiting\/} our notion of a `set'. We can no longer capture the
full, informal, notion of a set permitted by Cantor's na\"{\i}ve set
theory. However, our formal axiomatic set theories don't have any
paradoxes. This leaves an open question: how close to the na\"{\i}ve
notion of a set can we get using only formal (preferably axiomatic)
theories? 

\medskip

Prominent philosophers of mathematics such as Rudolf Carnap, and
mathematicians like Haskell Curry, have argued that the situation
depicted  in Figure~\ref{figure:ThirtyStruct} is only
temporary~\cite{Carnap:Logicist,Curry:Remarks}. They argue that one day
we will find a way of dealing with previously unsolvable informal (and
formal) problems using only axiomatic formal theory.

We will not deal here in detail with the objections to an axiomatic
foundation of mathematics; only noting in passing that the axiomatic
view is not uncontested~\cite{Berkeley:Treatise, Heyting:Intuitionist}.
Moreover, since the 1930s the boundary of axiomatic formal theory has
not changed. It seems at least possible, then, for the older view of
formal theory to offer something in our search for a more general
foundation for computation, particularly given the immense strides taken
in our understanding of axiomatic formal theory and computational theory
since the 1930s.

So, rather than debate whether the axiomatic position is ultimately
correct, we will simply \emph{assume\/} the structure of mathematics
resembles the one shown in Figure~\ref{figure:ThirtyStruct}. If we can
find a hypercomputational theory within this formal structure, we will
have a useful starting point. Whether such a theory can ultimately be
placed entirely within the bounds of an \emph{axiomatic\/} formal theory is
another argument: and one outside the bounds of this paper.

\section{A Metamathematical View of Computation}\label{s:StructECM}

\subsection{The Place of the Metatheory}

\begin{figure}
 \centering
      \subfigure[Moving from the structure of
      mathematics\lips\label{subfig:ThirtyZoom}]{\includegraphics[width=45mm,keepaspectratio]{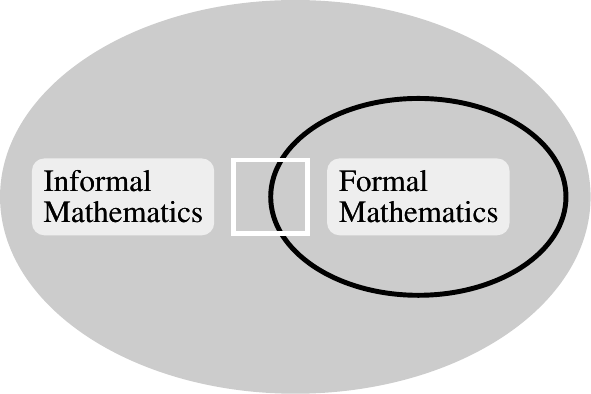}}
      \hspace{10mm} \subfigure[\lips\ to the structure of the
      metatheory\label{subfig:BasicOracle}]{\includegraphics[width=45mm]{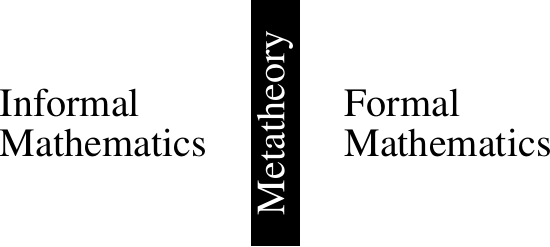}}

 \caption{\label{figure:PMeta}The Place of the Metatheory in Formal and
 Informal Mathematics}
\end{figure}

We will begin our investigation of computation by simplifying the
previous discussion somewhat, in order to focus more clearly on the
metatheory. Once we have the basic structure in place, we can use known
results of formal theory to give us an insight into the \emph{formal\/}
restrictions on the metatheory. Finally we will generalise these
restrictions using informal assumptions, leaving us with the foundations
of hypocomputational theories (the next section,
\S\ref{s:Hypocomputation} will explore these hypocomputational theories
in more detail).

Our starting point, therefore, is the simplified structure of
mathematics shown in Figure~\ref{figure:ThirtyStruct}. We are not
interested in any particular formal (or informal) theory, so we can
focus on an arbitrary region on the border between formal and informal
theory (shown in Figure~\ref{subfig:ThirtyZoom}). Restricting our focus
to this arbitrary region, we can make a further simplification to
produce the general structure shown in Figure~\ref{subfig:BasicOracle}.

The general structure shown in Figure~\ref{figure:PMeta} may seem
entirely too arbitrary to have any use. Its strength, though, comes from
precisely the same arguments used by Tarski to denounce metamathematics.
For we know from these arguments that \emph{under certain conditions\/}
the formal theory is independent of the metatheory. Under these specific
conditions the structural assumptions of the formal theory and the 
metatheory are said to be equivalent. Thus under these conditions we can 
use this structural equivalence to freely interchange both theory and
metatheory.

Thus by understanding these conditions, we can establish one aspect of
the relationship between the formal theory and the metatheory. We can go
further though, by studying the conditions where we \emph{cannot\/}
interchange the metatheory and formal theory, i.e.\ those conditions
where the formal theory is \emph{dependent\/} on the metatheory.

We will return to the relationship between this wider formal theory and
computation shortly (in \S\ref{ss:GenComp}). First, though, we will
study in more detail the conditions where the formal theory becomes
dependent on the metatheory for its own structure and behaviour.

\subsection{Formal Restrictions on the Metatheory}\label{ss:FormRes}

The greatest strength of any formal theory lies in the minimal number of
assumptions needed about the form or structure of a theory. This
strength also allows us to focus narrowly on a few basic assumptions,
leaving aside the considerations of any particular formal theory. From
this narrow focus, we can then return to a more general investigation of
the relationship between formal theory and metatheory.

\medskip

One feature of axiomatic formal theories is their association with
elementary number theory, in particular certain relations with the
\emph{ordinal numbers\/}\footnote{Strictly this association is only true
of all formal theories if Church's Thesis (discussed in
\S\ref{sss:WhyMeta}) is true. Although currently unproven, the evidence
for the thesis is strong (at least for axiomatic formal
theories)~\cite{Kleene:Mathematical, Kleene:Introduction}. We will
assume its truth for the remainder of the paper, although our conclusion
can be shown to hold even if Church's Thesis is false.}. Ordinal, or
`counting', numbers are simply the positive integers  (often written as
the set $\mathbb{Z}^+$ in older metamathematical literature, or
$\mathbb{N}^+$ in modern notation): $0$, $1$, $2$, $3$, $\ldots$.

From John von Neumann's work on set theory, we know we can obtain the
ordinal numbers from the most basic of sets: the empty set,
$\varnothing$\footnote{The following definition of ordinal numbers is
known to apply in both Georg Cantor's na\"{\i}ve set theory and modern
axiomatic theory. For a discussion of later work on axiomatic set theory
using this definition of ordinals see von Nuemann's own definition of an
axiomatic set theory~\cite{Neumann:Axiomatization}, and Jean van
Heijenoort's commentary on von Nuemann's 1923 paper~\cite[pp.
346--347]{Heijenoort:Frege}.

For convenience we will use
the usual modern notation for the empty set ($\varnothing$), rather than
the previously common notation used by von Neumann (O).
}. In all versions of set theory the empty
set is truly empty, having by definition no objects or elements as
members. Yet it still has form; enough, at least to define itself and the
ordinal numbers.

We can use this form to obtain the ordinal number as
follows~\cite{Neumann:Introduction}. First we start
with the empty set itself, and define this to be equal to the ordinal
$0$

$$0 = \varnothing,$$

we can then obtain the next ordinal by defining it as the set including
the previous ordinal. Since $\varnothing$ is empty we can ignore it, and
thus we define $1$ as

$$1 = \lbrace\varnothing\rbrace,$$

By extending this set and continuing, we obtain

$$2 = \lbrace\varnothing, \lbrace\varnothing\rbrace\rbrace,$$ $$3 =
\lbrace\varnothing, \lbrace\varnothing\rbrace, \lbrace\varnothing,
\lbrace\varnothing\rbrace\rbrace\rbrace, \ldots$$

We know we can use this sequence of ordinals as the foundation of at
least some axiomatic theories~\cite{Kleene:Introduction}. But does this
sequence tell us anything more general about the relationship between
formal theory and the metatheory?

\medskip

Ideally a formal theory should be both self-contained, and
self-describing. In other words, we should be able to use all the tools
provided by the theory, without having to bring in statements or
assumptions from elsewhere (especially those borrowed from an informal
theory). This requires, firstly, that the ideal formal is
\emph{closed\/}; all the objects necessary to the theory can be
described by the theory, and no object outside the theory has any
validity within the theory. Secondly an ideal formal theory is
\emph{complete\/}; meaning the theory covers the valid relationships
between the objects of the theory exactly\footnote{That is, all valid
relationships between objects are covered by the theory; all invalid
relationships are excluded by the theory; and the relationships
permitted by one statement of the theory do not differ from those
permitted by any other statement of the theory.}

Closure and completeness may be the ideal of any formal theory, but most
known axiomatic formal theories are not \emph{simultaneously\/} closed
and complete~\cite{Podnieks:Around}. Note that closure and completeness
are binary properties in an axiomatic formal theory --- you can't have
an axiomatic theory that is ``almost complete'' or ``almost closed''.
Formal theories failing the closure requirement we call \emph{open}, and
those failing the completeness requirements are called
\emph{incomplete}.

\medskip

Returning to the sequence of ordinals, we see that each member of the
sequence past $0$ contains at least two empty sets. For instance, the
sequence representing the ordinal $1$ ($ = \varnothing,
\lbrace\varnothing\rbrace, = \lbrace\varnothing\rbrace,$)
contains two empty sets:  $\varnothing$ ($ = 0$), followed by a
successor set containing only the empty set
($\lbrace\varnothing\rbrace$). Whether these two empty sets are in any
way `equivalent' is irrelevant. In practise, most formal theories do
assume the empty set is always equivalent to itself ($\varnothing \equiv
\varnothing$). Even so, the question in this context is meaningless: any
content of the empty set is unimportant, for the sequence depends only
on the ability to relate to the empty set.

What this sequence of ordinals does show, however, is that we can divide
a literal void into a series of non-interacting regions. In other words,
we can start with `nothing' and derive a minimal series of structural
assumptions. We can then relate these structural assumptions
to the closure assumptions of some formal theories.
Critically, though, these \emph{derived\/} structural assumptions should
not have to specify exactly what the \emph{actual\/} structure of the
void is. Questions about the void remain meaningless in the closed formal
theory.

Even from these minimal assumptions, though, we know that we can relate
non-interacting regions in some manner. We can actually be more specific, 
and infer this relationship as completeness in some formal theories.
Again, though, these relationships can not (and do not) say anything about the
\emph{actual\/} relationships permitted in the void. The void remains devoid of
structure and meaning in closed and complete formal theories.

\medskip

But what about open and incomplete formal theories?

\medskip

We can incorporate these theories into the same structure by removing
the minimal form of the empty set. This leaves us with a simple,
undifferentiated void; one entirely featureless and without any assumed
structure. Instead of a formal theory relying on itself to determine the
structure and relationships within this void, it must instead rely on
the metatheory to create these structures and relationships on behalf of
the formal theory.

\medskip

Let us therefore allow the metatheory to make a division in this void,
creating a region to the `left' and to the `right' of the division. What
requirements would a \emph{closed\/} formal theory have for this
division?

The first requirement would be the ability to differentiate between the
two regions. A void with a division is different to the original void,
and each region has its own identity. For instance, in the ordinal
sequence our first division creates two regions. One region remains the
void, the other is distinct enough to be called the empty set
($\varnothing$). Identity, though, is not quite enough. As we saw above,
no region can interact with the larger void. For example, in our
derivation of $1$ we obtained the set containing the empty set
($\lbrace\varnothing\rbrace$). But this same set appears in the
derivation of $2$: the new sets grows by adding further divisions, not
by coalescing previous divisions.

If our metatheory failed to uphold these structural criteria, we could not
call the resultant formal theory closed. For the theory now depends
\emph{directly} on the structural properties of the void --- properties
the theory can neither fully capture nor give meaning to.

However, we could call these `failed' theories \emph{open\/} and continue 
to use them. For open formal theories do not
require distinct identities for the region, nor do they exclude all
interactions between the divisions.

\medskip For the sake of argument we could then get the metatheory to
make another division, creating a new non-interacting region. We now
have two non-interacting regions; both separate from the larger void.
Since they are non-interacting, we cannot look to the void for
inspiration. We can, however, turn to the metatheory. What if the
metatheory could tell us how to relate the two regions?

In the case of the ordinals we have assumed the metatheory will do
exactly that. For instance, if we have the empty set ($\varnothing$) and
our next division produces the set containing the empty set
($\lbrace\varnothing\rbrace$), we know we can relate these two divisions
to produce $\varnothing, \lbrace\varnothing\rbrace$. Simplifying we
produce $\lbrace\varnothing\rbrace$, which our metatheory tells us is
the ordinal $1$.

What, though, if our metatheory cannot tell us how (or even if) two
regions are related? Again, we could not use such a metatheory to create
a \emph{complete\/} formal theory. We could, though, still use the
metatheory in \emph{incomplete\/} formal theories; accepting that under
some circumstances our metatheory would fail to guide us and that our
formal theories might therefore be inconsistent.

\medskip

Thus by studying the relationship between the metatheory and formal
theories, we can gain insight into the more general structure and
requirements of formal theories. However, we are not interested in
formal theory as such, but specifically in computational theory. And
computational theories have a few additional structural assumptions to
guide us. Of these assumptions, the most important is the
\emph{validity\/} of any conclusion (``output'') is independent of the
validity of the question (``input'')\footnote{In computational theory,
validity and truth are distinct (if related) concepts. Validity simply
means we have some expression of the input and output in terms of the
structural assumptions of the theory; for instance, a closed alphabet of
symbols. If we accept the input (or output) as valid, we still have
questions relating to the \emph{truth} of the statement. For example,
the statement ``$1 + 1 = 3$'' is clearly valid if we accept a closed
alphabet containing the symbols `$1$', '$3$', `$+$', `$=$'. Nonetheless,
we would have difficulty in creating an arithmetic theory expressing
the truth of this statement, assuming we accept the usual intuitive
notions and meanings of arithmetical statements. Loosely, then,
computational theories are essentially theories that try to determine
the truth (or falsity) of any statement valid in a closed theory.}. So do
the structures outlined in this section bear any relationship to the
structure of computational theory?

\subsection{A General Scheme for Computation}\label{ss:GenComp}

We saw in the previous section how we could infer a general relationship
between the metatheory and formal theories. But we have not yet shown
how this general structure relates specifically to computational
theories. In this section we will briefly outline this relationship,
before moving onto specific forms of computation in the following
sections.

\medskip

In relating the general structure of formal theories to computational
theories, we will start with Alan Turing's consideration of the ordinal
sequence, published in his 1939 paper \emph{Systems of Logic Based on
Ordinals\/}~\cite{Turing:Systems}. In \emph{Systems of Logic\/} Turing
builds on certain results in number theory studied in his earlier 1936
paper, \emph{On Computable Numbers, With an Application to the
\emph{Entscheidungsproblem}\/}~\cite{Turing:Computable}. We are  less
interested here in the results of \emph{Systems of Logic}, than in
Turing's discussion of his \emph{a\/}-machines, first defined in
\emph{On Computable Numbers\/} (and briefly earlier discussed in
\S\ref{ss:RetToMet}). Again the details of the \emph{a\/}-machines are
unimportant here; for the moment we will simply note that
\emph{a\/}-machines are accepted as the foundation of at least one
computational theory~\cite{Gandy:Confluence}\footnote{We will return to
the details of Turing's \emph{a\/}-machines when we study strictly
computational theories in more detail in \S\ref{s:Computation}}.

In \emph{On Computable Numbers\/} Turing focuses on the decision problem
(in German the \emph{Entscheidungsproblem\/}) in elementary arithmetic.
This decision problem relates to Hilbert's program of formalising
mathematics, and is essentially the question of whether we can always
determine whether a statement in elementary arithmetic is true or false.
Turing redefines this problem as the actions of an \emph{a\/}-machine,
showing that \emph{in general\/} we cannot construct an
\emph{a\/}-machine to answer the decision problem. Since we cannot
construct a general \emph{a\/}-machine to answer the decision problem,
by extension Turing concludes the decision problem is undecidable in
elementary arithmetic.

\medskip

Ultimately, only two answers to the decision problem exist: either the statement is
true, or it is false. What, then, if we had some other way of answering
the decision problem for an arbitrary statement in arithmetic? A way
that could not be expressed in terms of computational theory, but which
was still capable of giving a valid answer? 

Once answered, the question could then be \emph{used\/} in a theory of
arithmetic. The problem is that we cannot construct an axiomatic theory
to both answer the question \emph{and\/} make use of the answer.

Could we, however, construct two different theories? One theory could
answer the question for us: although obviously this theory could not be
constructed in the same manner as a theory of arithmetic. We could,
though, construct a modified axiomatic theory of arithmetic making use
of the first theory, allowing us to answer the general decision problem
in this new, modified theory

Since \emph{a\/}-machines cannot in general answer the decision problem,
this would still not lead to a \emph{full\/} axiomatic theory. Nor could
we express the new theory simply in terms of the actions of an
\emph{a\/}-machine. But perhaps we could create a new type of machine,
allowing us to separate the two theories. This would then allow us at
least to state \emph{part\/} of the theory in terms of the actions of an
\emph{a\/}-machine (i.e.\ axiomatically).

\medskip

This argument is essentially the only one used by Turing in
\emph{Systems of Logic}, although he also concluded that we could create
these hybrid theories, using a modified \emph{a\/}-machine~\cite[pp.
172--173]{Turing:Systems}:

\begin{quotation}

Let us, therefore, suppose that we are supplied with some unspecified
means of solving number-theoretic problems; a kind of oracle as it were.
We shall not go any further into the nature of this oracle apart from
saying that it cannot be a machine. With the help of the oracle we could
then form a new kind of machine (call them \emph{o\/}-machines), having
as one of its fundamental processes that of solving a given
number-theoretic problem.

\end{quotation}

Note that oracle machines \emph{are not\/} intrinsically any more
powerful than \emph{a\/}-machines~\cite[\S 4, p. 173]{Turing:Systems}.
Within an axiomatic theory, the actions of both types of machines are
exactly equivalent~\cite[Chapter 24]{Atallah:Algorithms}. The only
difference between \emph{a\/}-machines and \emph{o\/}-machines is the
reliance on the metatheory. For \emph{a\/}-machines their behaviour is
independent of the metatheory, since the axiomatic theory governing
their behaviour and the metatheory are deemed to be equivalent. Under
computational assumptions, these conclusions also hold for
\emph{c\/}-machines; since we are assuming the structural theories
underpinning \emph{a\/}-machines and \emph{c\/}-machines are equivalent.

By contrast, the \emph{general\/} behaviour of \emph{o\/}-machines can only be 
specified by reference to both the axiomatic theory and the metatheory.
Only in certain special cases can we fully specify the actions of the
\emph{o\/}-machines using only the axiomatic theories. Within this
special case, the actions of the \emph{o\/}-machines are equivalent to the
action of the \emph{a\/}-machines. But only within this special case can
we ignore the metatheory. For \emph{o\/}-machines in general, the
metatheory also acts as a filter, coercing the wider structures of the
formal theory into an axiomatic theory. This also leaves open the
possibility of constructing more general \emph{c\/}-machines, if we base
our choice mechanism on these wider formal structures.

\medskip

More general formal theory, then, can be used with axiomatic formal
theories. All we have to do is ensure that under certain conditions the
completeness and closure assumptions of the axiomatic formal theory can
be met. We can go further, since Turing arguments neatly divide the
requirements of the axiomatic formal theory from the requirements of the
general formal theory. We also know that this general formal theory can
allow computation; for instance computing the result of an arbitrary
decision problem. Using Turing's arguments, we can elaborate on the
general structure of formal theories outlined in \S\ref{ss:FormRes}, and
specifically on computation and computable theories.

\medskip

Let us start with Turing's \emph{o\/}-machines, as an exemplar for
computable theories. We know each \emph{o\/}-machine is essentially an
\emph{a\/}-machine, with equivalent properties. Since each
\emph{a\/}-machine has a unique description governing the behaviour of
the \emph{a\/}-machine~\cite{Turing:Computable}, \emph{o\/}-machines
also have such a description. The nature of this description is
unimportant, we only need to know that it exists. Turing's
\emph{o\/}-machines (like \emph{a\/}-machines) move stepwise through the
description, and for the moment we will also adopt that restriction.

\medskip

For the sake of argument, we will assume the \emph{o\/}-machine has
reached a certain point and that a next move is possible. For an
\emph{a\/}-machine the next move can be deduced simply by applying the
axioms to the current machine description. But for an \emph{o\/}-machine
this might not be the case: the next move may require consultation of
the metatheory (via the oracle). In this case the next move for the
\emph{o\/}-machine exists in a featureless, undifferentiated void; about
which the axioms of the \emph{o\/}-machine can say nothing. We will call
this void the \emph{machine void}, since the nature of the void depends
partly on the machine (as we shall soon see).

\begin{figure}
 \centering
   \subfigure[The Structure of a Formal
   Theory\label{figure:VoidLayout}]{\includegraphics{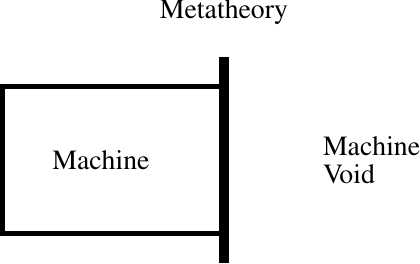}}
   \hspace{10mm} \subfigure[\label{figure:MechView}The Machine View of a
   Formal Theory]{\includegraphics{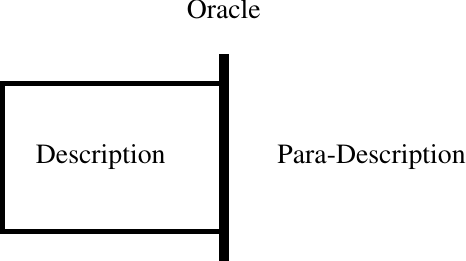}}
 \caption{\label{figure:GenComp}The General Structure of Computation}
\end{figure}

Nonetheless, an \emph{o\/}-machine can access this void via the
metatheory --- if the metatheory allows it. The metatheory, therefore,
acts as a barrier between the \emph{o\/}-machine and the void. Together,
these arguments produce a general structure along the lines of
Figure~\ref{figure:VoidLayout}.

From the perspective of the \emph{o\/}-machine, the metatheory acts as
an oracle. Note, however, that the \emph{o\/}-machine cannot see `past'
the oracle: the void is not part of the description of the
\emph{o\/}-machine. Nevertheless, the \emph{o\/}-machine acts \emph{as
though\/} the description \emph{did\/} extend `past' the oracle. It can
do this because it can always apply the machine axioms to the current
machine configuration. But this can be no more than a projection of the
axioms: the \emph{o\/}-machine has no guarantees over the
\emph{actual\/} structure and behaviour of the projection. We will,
therefore, call this projection the \emph{para-description\/} of the
machine, since it is related to (but not the same as) the machine
description. Thus the \emph{o\/}-machine's view of the formal theory
shown in Figure~\ref{figure:VoidLayout} looks more like
Figure~\ref{figure:MechView}.

\medskip

For an \emph{o\/}-machine describing a complete and closed theory, 
the metatheory must constrain the view of
the void presented by the metatheory to the machine. Although we have
argued for the structure in Figure~\ref{figure:GenComp} from the actions
of \emph{o\/}-machines, we claim that this structure is applicable to
both axiomatic and more general formal theories. We make this claim
because we can relax the completeness and closure requirements of
\emph{o\/}-machines, while still creating a formal theory under the
control of the metatheory. Some of the machines created using this
general formal theory will be equivalent to axiomatic formal theories of
computation, such as the on used by \emph{a\/}-machines. But other
models of computation will be more general than strictly axiomatic
theories allow, as we will explore in the following sections.

\section{Hypocomputation}\label{s:Hypocomputation}

Consider an \emph{o\/}-machine mimicking the actions of an
\emph{a\/}-machine. The para-description for the \emph{o\/}-machine
would obey the same rules as the machine description, and would be
structurally equivalent to the machine description. We could, though,
imagine an \emph{o\/}-machine at the other extreme, where the
\emph{o\/}-machine consults the oracle for every move. This
\emph{o\/}-machine would effectively have no para-description, since the
behaviour of the machine is governed by the oracle and not the axioms of
the machine\footnote{In between these extremes a number of possibilities
exist, usually referred to as the \emph{arithmetic hierarchy}. In an
unpublished paper of the author, we show that this same basic scheme can
be used to describe all members of the arithmetic
hierarchy~\cite{Love:Computation}.}.

Nonetheless, if the metatheory is constructed so that the oracle obeys
the completeness and closure rules of the machine description, the
\emph{o\/}-machine would act in every way as expected. Thus the simple
presence or absence of the para-description does not affect the
behaviour of the \emph{o\/}-machine. However, this type of
\emph{o\/}-machine is completely reliant on the \emph{metatheory\/}
presenting a para-description (through the oracle) in a manner
consistent with the machine description.

But what if the metatheory \emph{does not\/} present the
para-description in a manner consistent with the machine description?
Obviously if the metatheory fails to uphold \emph{both\/} the
completeness and closure assumptions of the machine description, the
resultant machine is unlikely to resemble an \emph{o\/}-machine (or be
anywhere near as useful). Nonetheless, we don't have to drop both
assumptions. We could allow the metatheory to preserve \emph{either\/}
the completeness \emph{or\/} the closure assumptions of the machine
description.

We then enter the realm of computation we call \emph{hypocomputation},
since we obtain machines similar to, but less powerful than,
\emph{o\/}-machines. If we allow the metatheory to preserve only the
completeness assumptions of the machine description in the
para-description, we obtain machines we will call \emph{complete
hypocomputers}. Likewise, if we allow the metatheory to preserve only
the closure assumptions of the machine description in the
para-description, we obtain \emph{closed hypocomputers}.

\medskip

The existence of a \emph{partial\/} para-description is the key feature
of a hypocomputer. For this partial para-description allows us to make
some assumptions about the machine void, and as a consequence to reason
about \emph{some\/} aspects of the machines behaviour. Crucially,
however, we cannot reason about the \emph{entire\/} behaviour of the
machine, knowing only the machine description, or the machine's current
configuration (as we can for computational machines). But why should we
even consider such machines?

\medskip

From a theoretical perspective, hypocomputers are extremely useful,
given the relative scarcity of complete and closed formal theories. For
example, mathematical research during the late 19\th\ century uncovered
a large number of useful formal theories containing
paradoxes~\cite[Chapter III, \S11]{Kleene:Introduction}. Indeed David
Hilbert started his  metamathematical program to find a paradox free
foundation for mathematics~\cite{Detlefsen:Hilberts}.

Modern developments in mathematics have largely followed Hilbert in
forbidding paradoxes in formal theories. For instance Georg Cantor's
(na\"{\i}ve) set theory contains a number of interesting paradoxes,
which modern axiomatic set theories carefully step around~\cite[Chapter
III, \S12]{Kleene:Introduction}.

While computable theories can explore these modern, axiomatic theories
(as we saw in \S\ref{s:Introduction}), the larger realm of formal
theories remains out of reach. Moreover, developments around Church's Thesis
suggest we will never be able to reach this region of formal theory, if we insist on
using only axiomatic theories. However, we can explore these realms
using the tool of hypocomputational formal theories, whose actions are
naturally paradoxical.

\medskip

Even if we are not interested in theories containing paradoxes,
hypocomputation has many practical uses. No modern computational machine
truly matches the assumptions of Turing's \emph{a\/}-machines. Instead
we have learned to build ``almost'' closed and complete devices whose
behaviour is strongly reminiscent of an \emph{a\/}-machine.

As we said before, though, no axiomatic formal theory can be ``almost
complete'' or ``almost closed''. Either the theory is complete or it is
not. Likewise with closure.

At present, however, we are studying and defining the actions of modern
computational machines using only axiomatic formal theories. But this
requires hiding the assumptions that ``almost'' match the formal axioms
of the theory with the machine description, making it hard to reason
about the machine's behaviour~\cite{Glass:Defence}. If we focused
instead on hypocomputational formal theories, we might be able to reason
more flexibly about the machine behaviour: as long as we don't expect
perfection.

\medskip

Hypocomputers are thus an unexplored area of formal theory with many
interesting theoretical and practical possibilities. Moreover, we
already have devices resembling hypocomputers to study, making this
possibly the most accessible realm of computation.

\section{Computation}\label{s:Computation}

More traditional axiomatic computable theories have been well studied
and described by others~\cite{Atallah:Algorithms, Herken:Universal,
Kleene:Mathematical}. We have also stated how our general picture of
formal theories relates specifically to these computational theories  in
\S\ref{ss:GenComp} and in the introduction to \S\ref{s:Hypocomputation},
so we will consider only the barest details here.

\medskip

Turing defined his \emph{a\/}-machines by considering the actions of an
idealised human computer, concluding that~\cite[p.
231]{Turing:Computable}:

\begin{quotation}

We may compare a man in the process of computing a real number to a
machine which is only capable of a finite number of conditions $q_1$,
$q_2$, $\ldots$, $q_R$ which will be called
``\emph{m\/}-configurations''. The machine is supplied with a ``tape'',
(the analogue of paper) running through it, and divided into sections
(called ``squares'') each capable of bearing a ``symbol''. At any moment
there is just one square, say the \emph{r\/}-th, bearing the symbol
$\mathfrak{S}(r)$ which is ``in the machine''. We may call this square
the ``scanned square''. The symbol on the scanned square may be called
the ``scanned symbol''. The ``scanned symbol'' is the only one of which
the machine is, so to speak, ``directly aware''. However, by altering
its \emph{m\/}-configuration the machine can effectively remember some
of the symbols which it has ``seen'' (scanned) previously. The possible
behaviour of the machine at any moment is determined by the
\emph{m\/}-configuration $q_n$ and the scanned symbol $\mathfrak{S}(r)$.
This pair $q_n$, $\mathfrak{S}(r)$ will be called the ``configuration'':
thus the configuration determines the possible behaviour of the machine.
In some of the configurations in which the scanned square is blank
(i.e.\ bears no symbol) the machine writes down a new symbol on the
scanned square: in other configurations it erases the scanned symbol.
The machine may also change the square which is being scanned, but only
by shifting it one place to right or 1eft.

\end{quotation}

\begin{figure}
   \centering \includegraphics{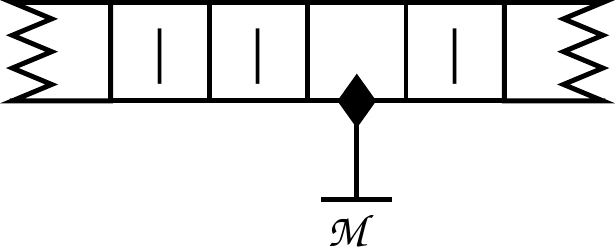}
   \caption{\label{figure:ExampleTM}A Sketch of a Turing
   \emph{a\/}-machine}
\end{figure}

These \emph{a\/}-machines may also be represented graphically, along the
lines of Figure~\ref{figure:ExampleTM}. We draw a tape divided into a
series of squares, each one either blank or containing a symbol. The
machine (and its configuration) we show as $\mathcal{M}$ in
Figure~\ref{figure:ExampleTM}, separated from the tape by the ``head''
of the machine. The only view of the tape from the machine is via this
head; more specifically the machine can only see the scanned symbol
(shown by the square marked with a $\blacklozenge$ in
Figure~\ref{figure:ExampleTM}).

Applying the same terminology used in the rest of the paper to
Figure~\ref{figure:ExampleTM}, $\mathcal{M}$ is the machine description
and the tape is the para-description of the machine. The tape thus
represents the machine's \emph{view\/} of the machine void. In an
\emph{a\/}-machine, the para-description is always consistent with the
machine void, so the distinction between the para-description and the
machine void can be ignored. In effect the machine has direct access to
the tape.

For an \emph{o\/}-machine, however, the view of the machine void is
mediated by the metatheory. Thus the read-head appears to the machine as
an oracle supplying the next symbol. Depending on the particular
construction of $\mathcal{M}$, an \emph{o\/}-machine may or may not have
a para-description for a particular move. An \emph{o\/}-machine's view
of the tape is thus governed by the metatheory (via the oracle), as we
would expect.  The differences between the assumptions of the
\emph{a}-machine and those of the \emph{o}-machine can be used to
produce a variety of computational automata. As the details of these
automata, and the distinctions between them, are well studied and
described~\cite{Hopcroft:Formal, Starke:Abstract, Trakhtenbrot:Finite},
we will not discuss them further in this paper.

\section{Hypercomputation}\label{s:Hypercomputation}

In general, then, the machine void (or ``tape'') is only viewed
\textbf{indirectly} by the machine. Instead, the machine only has direct
access to the para-description; effectively the tape as the machine
\emph{believes\/} it exists. Usually this distinction is irrelevant,
since axiomatic formal theory assumes the behaviour of the void is
specified by the para-description of the machine.

If more than one machine exists, though, the machine may either share a
tape (machine void), or each machine may have its own tape (machine
void). Axiomatic theories assume \emph{but cannot prove\/} these
situations are equivalent. In the general case, though, this assumption
breaks down; as the decision problem illustrates.

Hypercomputers resolve this problem by eliminating the possibility of
machines using different voids. This situation cannot be described
axiomatically, since it requires the machine configuration of one
machine to \emph{directly\/} alter the machine configuration of another.
Nonetheless, this action is permitted within the scope of formal
theories.

\medskip

\begin{figure}
 \centering
   \subfigure[Copy into a shared (overlapping) machine
   void\label{subfig:OverCopy}]{\includegraphics[width=45mm,keepaspectratio]{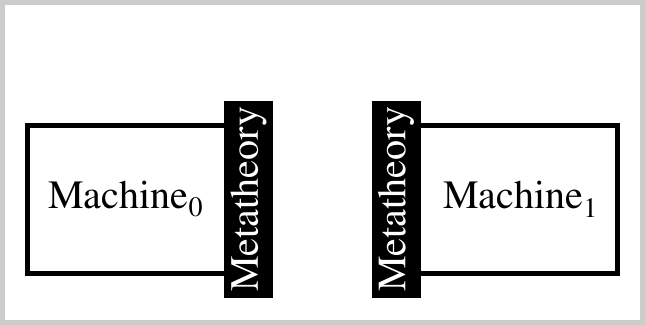}}
   \hspace{10mm} \subfigure[Copy into new (non-overlapping) machine
   voids\label{subfig:NewCopy}]{\includegraphics[width=45mm,keepaspectratio]{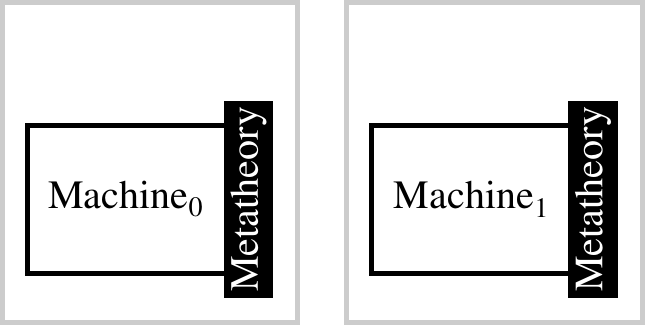}}
 \caption{\label{figure:FormCopy}Two Outcomes for a Copy in a Formal
 Theory}
\end{figure}

To explore this problem in more detail, consider a simple
\emph{a\/}-machine reading symbols from the ``tape'' and writing the
same symbol back to the ``tape''. In general, this machine is asking the
metatheory to supply the next symbol from the machine void, which the
machine then assumes is ``written back'' to the machine void in some
manner. However, since the machine cannot see the machine void directly,
the machine \emph{cannot\/} differentiate between the two situations
illustrated in Figure~\ref{figure:FormCopy}. The difference between the
situations becomes critical, though, when we try to relate machines;
especially when we use the behaviour of one machine to reason about
another.

\medskip

For example, consider a problem posed by Turing in \emph{On Computable
Numbers}. Just after Turing's proof that the decision problem is
undecidable for all \emph{a\/}-machines, Turing goes on to say~[p.
248]\cite{Turing:Computable}:

\begin{quotation}

We can show further that \emph{there can be no machine $\mathcal{E}$
which, when applied with the S.D $\llbracket($standard
description\footnote{Essentially, the standard description is the
canonical sequence of symbols uniquely describing the particular
\emph{a\/}-machine under study
(\cite[\S5]{Turing:Computable}).}$)\rrbracket$ of an arbitrary machine
$\mathcal{M}$, will determine whether $\mathcal{M}$ ever prints a given
symbol\/} (0 say).

\end{quotation}

This proof is obtained by breaking the stated problem in to a sequence
of sub-problems, each described by a particular machine~\cite[p.
248]{Turing:Computable}:

\begin{quotation}

We will first show that, if there is a machine $\mathcal{E}$, then there
is a general process for determining whether a given machine
$\mathcal{M}$ prints 0 infinitely often. Let $\mathcal{M}_1$  be a
machine which prints the same sequence as $\mathcal{M}$, except that in
the position where the first 0 printed by $\mathcal{M}$ stands,
$\mathcal{M}_1$ prints $\overline{0}$. $\mathcal{M}_2$ is to have the
first two symbols 0 replaced by $\overline{0}$, and so on. Thus, if
$\mathcal{M}$ were to print

$$\mathrm{\mathrm{A}} \mathrm{B} \mathrm{A} 0 1 \mathrm{A} \mathrm{A}
\mathrm{B} 0 0 1 0 \mathrm{A} \mathrm{B} \ldots $$

then $\mathcal{M_1}$ would print

$$A \mathrm{B} \mathrm{A} \overline{0} 1 \mathrm{A} \mathrm{A}
\mathrm{B} 0 0 1 0 \mathrm{A} \mathrm{B} \ldots $$

and $\mathcal{M_2}$ would print

$$A \mathrm{B} \mathrm{A} \overline{0} 1 \mathrm{A} \mathrm{A}
\mathrm{B} \overline{0} 0 1 0 \mathrm{A} \mathrm{B} \ldots  $$

\end{quotation}

Note how this sequence assumes the para-description of each machine is
shared, i.e.\ that the situation in Figure~\ref{subfig:OverCopy}
applies. For instance, we assume the input of $\mathcal{M}_1$ is not
only \emph{equivalent\/} to the output of $\mathcal{M}$ but \textbf{is}
the output of $\mathcal{M}$. We can preserve the equivalence between the
output of $\mathcal{M}$ and the input of $\mathcal{M}_1$ whether the
machine void is shared or not. But mere equivalence is not sufficient.
What we are claiming is that para-description of one machine
\emph{determines\/} the behaviour of another. But the machine
\emph{itself\/} \textbf{cannot} make this determination: only the
metatheory can.

\medskip

Moreover, if the machine voids are not overlapping, the metatheory will
be drawing the machine behaviour of the next machine from a
\emph{different\/} machine void. The metatheory \emph{may\/} be able to
mask the difference, but in general the machine will be unaware either
way.

We can see this from Turing's conclusion of the proof~\cite[p.
248]{Turing:Computable}:

\begin{quotation}

Now let $\mathcal{F}$ be a machine which, when supplied with the S.D of
$\mathcal{M}$, will write down successively the S.D of $\mathcal{M}$, of
$\mathcal{M}_1$, of $\mathcal{M}_2$, \lips\  (there is such a machine).
We combine $\mathcal{F}$ with $\mathcal{E}$ and obtain a new machine,
$\mathcal{G}$. In the motion of $\mathcal{G}$ first $\mathcal{F}$ is
used to write down the S.D of $\mathcal{M}$, and then $\mathcal{E}$
tests it, :0: is written if it is found that $\mathcal{M}$ never prints
0; then $\mathcal{F}$ writes the S.D of $\mathcal{M}_1$ and this is
tested, :0: being printed if and only if $\mathcal{M}_1$ never prints 0;
and so on. Now let us test $\mathcal{G}$ with $\mathcal{E}$. If it is
found that $\mathcal{G}$ never prints 0, then $\mathcal{M}$ prints 0
infinitely often; if $\mathcal{G}$ prints 0 sometimes, then
$\mathcal{M}$ does not print 0 infinitely often.

\end{quotation}

$\mathcal{M}$ only exists if a general solution to the decision
(halting) problem exists. But a general solution to the decision problem
is known to be impossible. Turing thus concludes $\mathcal{M}$ cannot
exist, and hence $\mathcal{E}$ cannot exist either~\cite[p.
248]{Turing:Computable}.

Our problem lies in the definition of the \emph{a\/}-machine
$\mathcal{F}$. This \emph{a\/}-machine can indeed create a succession of
\emph{a\/}-machine descriptions ($\mathcal{M}$, $\mathcal{M}_1$,
$\mathcal{M}_2$, \lips). However \emph{a\/}-machine $\mathcal{F}$ cannot
make any guarantees about the para-description for any of these
machines. We need this guarantee, though, to be able to reason about the
actions of any particular machine. For instance, to \emph{ensure\/} the
symbol :0: is only printed ``\emph{if and only if $\mathcal{M}_1$ never
prints 0\/}''.

\medskip

The only way to obtain this guarantee would be for the
\emph{metatheory\/} to ensure all machine voids overlap. Axiomatic
theories, though, cannot themselves provide this guarantee (otherwise
they could solve the decision problem). If the metatheory can make this
guarantee, the machine becomes not computational but
\emph{hypercomputational}.

A hypercomputer, then, is simply a computational machine with a
guarantee that machine voids are \emph{always\/} shared. Only under
these conditions can we use the para-description to reason about the
action of the machine, as though the metatheory did not exist.

\section{Physical Instantiation}

Before we conclude, we will briefly examine a question often asked in
the hypercomputation literature: ``\emph{can we actually \emph{build\/}
a hypercomputer\/}''?

Answering this question successfully requires us to address two distinct
problems relating to a machine's void. In essence, our problem is
finding a way of preserving the illusion that a particular machine's
projection serves as a model of reality. This requires, firstly, that we
can ensure the behaviour of the machine's void matches the projection of
that machine, by, for instance, ensuring the assumed `inputs' to the machine do
not violate the closure assumptions of that particular machine.
Secondly, we must find some way of constraining the behaviour of the
machine's void; but without requiring physically impossible means to do
it.

\medskip

These problems may seem trivial, and indeed are often treated simply as
`implementation details'. Nonetheless, if our experience with
computational machines is a reliable guide, neither condition is
possible to satisfy fully. Instead we have only been able to build
``almost computational'' machines: learning instead to live with the
consequences.

For example, we may decide to filter the inputs to the machine to meet
the first constraint. Any input value higher (or lower) than the machine
will accept is simply held to the maximum (or minimum). This behaviour
is common in control systems --- however it is not sufficient to fully
meet the second condition. What we need is not a means of ensuring the
input \emph{does not\/} exceed the maximum value, but that the input
\textbf{can not} do so. What happens if the filter itself malfunctions
and enters an incorrect value into the system? If the filter simply
truncates a value to fit the input criterion, is this acceptable: or
have we just introduced an unexpected new behaviour into the machines
para-description?

In real systems, we must find some way of addressing these questions.
Control systems, for example, often have as many behaviours dedicated to
the detection of `errors', as they do for actual control. Yet we
\emph{have\/} been able to build computational control systems.
Moreover, such systems are in widespread use --- indeed many human lives
depend on them working as expected.

\medskip

Perhaps, then, the question should not be ``\emph{can we build a
hypercomputer\/}''. The simple answer to that question would appear to
be \emph{no}. But that still leaves open the question: ``\emph{how close
can we get to building a hypercomputer\/}''?

In answering that question we have nearly 60 years experience with
``almost computers'' to guide us. So let us briefly re-examine the two
problems, raised earlier, using this experience as a guide to question
of whether we can build an ``almost hypercomputer''.

\subsection{Projecting Into the Void}

The \emph{expected\/} behaviour of a machine's void is, to a large
extent, governed by the design of that machine. Similarly, the
\emph{actual\/} behaviour of the machine's void can be left to the
implementation of the machine. Conveniently, separating the `design'
phase from the `implementation' phase has been a longstanding principle
in the implementation of computational machines. We will, therefore,
follow tradition and focus on the design of a hypercomputer in this
section, leaving the discussion of the implementation for the next
section, \S\ref{ss:ConstrainVoid}.

\medskip

\begin{table}
  \centering

  \begin{tabular}{p{3cm}ccc}

   \toprule%

   \begin{minipage}{3cm}
      \begin{center}
         \textbf{Research Strand}
      \end{center}
   \end{minipage} &

  \begin{minipage}{2.5cm}
      \begin{center}
      \end{center}
   \end{minipage} &

   \begin{minipage}{2.5cm}
      \begin{center}
         \textbf{Is the Description Closed?}
      \end{center}
   \end{minipage} \\

   \midrule%

   Formal & Yes & Yes \\

   Semi-formal & Yes & No \\

   Object-Oriented & No & Yes \\

   Holistic & No & No \\

   \bottomrule%

   \end{tabular}

   \medskip

   \caption{\label{table:SoftDiv}Assumptions of Software Descriptions by
   Research Strand~\cite[Table 3.1]{King:Software}}

\end{table}

Researchers have been actively studying design methods for digital
computation machines since the late 1940s~\cite{Avison:Information}.
During this time, many hundreds of methods have been proposed, giving a
potentially vast pool of methods to consider. Happily, though, all the
methods in this pool rest on only a few basic assumptions about the
nature of the software description~\cite{King:Software}. We can
summarise these assumptions in Table~\ref{table:SoftDiv}.

Of these categories, only design methods from the formal strand assume a
complete and closed description. Given the reliance of hypercomputation
on complete and closed description, it seems safe to assume we could
reuse these methods for the design of a hypercomputer. But could we go
further?

\medskip

We know from experience with strictly computational machines, that
incompleteness and openness in the wider software description can (an
indeed must) be eliminated from the executable program description. The
problem in conventional software design, though, is achieving this
elimination without compromising the intent of the software
description~\cite[Chapter 9]{King:Software}.

Hypercomputation, though, removes one of the main problems with formal
design methods, namely the inability to separate actions that create new
machine voids from actions that copy into the same machine void. This,
in turn, makes it considerably easier to cleanly separate software
assumptions from program assumptions, easing the entire design process.

\medskip

For example, consider one of the oldest design principles used by many
conventional software and program design methods: coupling and cohesion,
first defined by Glenford Myers and Larry Constantine in the early
1970s. Broadly, ``\emph{coupling is a measure of how connected two items
are, and cohesion is a measure of how much something makes
sense\/}''~\cite{Ambler:Process}. While this definition maybe vague,
many researchers have attempted to make use of these metrics, leading to
the `definition' of cohesion shown in Table~\ref{table:Cohesion}.

\begin{table}

   \begin{center}

      \begin{tabular}{p{2.5cm}p{4.5cm}p{2cm}}

         \toprule%

         \begin{minipage}[c]{2.5cm}
            {
               \begin{center} \textbf{Level Name} \end{center}
            }
         \end{minipage} &

         \begin{minipage}[c]{4.5cm}
            {
               \begin{center} \textbf{Level Description} \end{center}
            }
         \end{minipage} &

         \begin{minipage}[c]{2cm}
            {
               \begin{center} \textbf{Cohesion Level} \end{center}
            }
         \end{minipage}\\

         \midrule%

         \emph{Object\/} & Each operation provides functionality which
         allows object attributes to be modified or inspected & Very
         High \\

         \emph{Functional\/} & Each part of the module is necessary for
         the execution of a single function & Very High \\

         \emph{Sequential\/} & The output from element in the module
         serves as the input for some other module & High \\

         \emph{Communicational\/} & All of the elements of a module
         operate on the same input data or produce the same output data
         & Medium \\

         \emph{Procedural\/} & All elements of a module describe a
         single control sequence & Medium \\

         \emph{Temporal\/} & All modules that are activated at a single
         time are grouped together & Medium \\

         \emph{Logical\/} & Modules that perform similar functions are
         put together into a single module & Low \\

         \emph{Coincidental\/} & Parts of a module are not related by
         simply bundled into a single module & Very Low \\

         \bottomrule%

      \end{tabular}

      \caption{\label{table:Cohesion}Levels of
      Cohesion~\cite{Myers:Reliable,Page-Jones:Practical,Sommerville:Engineering,Stevens:Structured}}

   \end{center}

\end{table}

While cohesion may be popular in both program and software design
methods, we can easily show cohesion is impossible to define in
axiomatic formal theories. As a metric, cohesion therefore makes little
sense in the design of programs for conventional computers.

\begin{figure}
   \begin{center}
      \subfigure[Design
      1\label{subfig:CCP1}]{\includegraphics{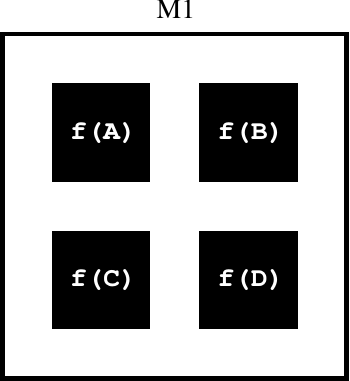}}
      \hspace{2cm} \subfigure[Design
      2\label{subfig:CCP2}]{\includegraphics{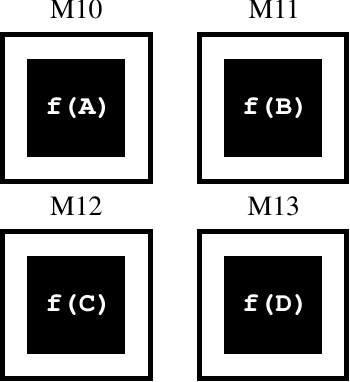}}
      \caption{Two Equivalent Module Designs}\label{figure:CohModDesign}
   \end{center}
\end{figure}

For instance, consider the description of the four self-contained
functions shown in Figure~\ref{figure:CohModDesign}; function
\texttt{A}, function \texttt{B}, function \texttt{C} and function
\texttt{D}. For simplicity's sake, each of these functions only relies
on itself for its definition, and  each function is unrelated in intent
or purpose. Design $1$  in Figure~\ref{subfig:CCP1} shows all four
functions in the same module, \textsc{M1}. Therefore, Module \textsc{M1}
has coincidental cohesion (following Table~\ref{table:Cohesion}); the
lowest level of cohesion. Since each function is unrelated, we should
separate each function into its own module, producing design $2$ shown
in Figure~\ref{subfig:CCP2}. According to Table~\ref{table:Cohesion},
this design now has the highest level of cohesion, functional cohesion.

As program descriptions, the two designs in
Figure~\ref{figure:CohModDesign} are functionally equivalent. It is a
trivial task to map one to the other. However, the design in
Figure~\ref{subfig:CCP1} has only one machine void, whereas the design
in Figure~\ref{subfig:CCP2} has \emph{between\/} one and four. That is,
all the machines might be sharing a common void, or each machine might
be interfacing to a distinct void. Or we might be in a situation
in-between these extremes. But under no circumstances can the axiomatic 
theories tell us which situation we are dealing with. So while
the designs in Figure~\ref{figure:CohModDesign} are functionally
equivalent, the implemented behaviour of the two design may be very
different.

In the design of conventional computers, solving this problem is
non-trivial (and indeed impossible in general). For a hypercomputer,
though, we know the creation of new machine voids is forbidden. If
Figure~\ref{figure:CohModDesign} referred to hypercomputational design,
then we know that each design has exactly one machine void. This makes
it much easier to reason about the machine behaviour, since changes in
the designed behaviour can be more easily isolated from the implemented
behaviour.

\subsection{Constraining the Void}\label{ss:ConstrainVoid}

While reasoning about hypercomputational design may be easier,
implementing hypercomputers is likely to prove much more difficult than
conventional computers. For the same properties that favour the design
process, work against us in the implementation phase. By forbidding
overlapping voids \emph{every\/} part of the hypercomputational
description is exposed to \emph{all\/} changes to the common machine
void. This means the behaviour of the void is far more critical, because
at no point can the behaviour of the void violate the closure and
completeness assumptions of \emph{any\/} part of the machine
description.

\medskip

Currently, this point is poorly understood in many proposed physical
implementations of hypercomputers. For conventional computers, we can
use non-overlapping machines to contain violations of the closure and
completeness assumptions. Some physical implementations even exploit
this separation of machine voids, easing the implementations of the
machine.

However, this approach is completely invalid for hypercomputational
machines, since we cannot isolate a single machine void from any other.
Even so, most proposed hypercomputers follow a split model similar to
Turing \emph{o\/}-machines discussed in \S\ref{ss:GenComp}. But
hypercomputers cannot be \emph{o\/}-machines: \emph{both\/} sides of the
split use the same machine void, with the same completeness and closure
assumptions.

\medskip

For example, a particularly problematic (though common) implementation
of a hypercomputer involves signalling the result of a computation to a
conventional computer. Such a scheme requires not only that each part
obey the completeness and closure assumption, \emph{but every
\textbf{possible} path\/} between the two parts. What happens if one
part fails and sends/receives the signal incorrectly? Can we guarantee
\emph{absolutely\/} that no signal received \emph{only\/} occurs if no
signal is sent?

These are not simply `implementation details'. Our hypercomputer is
using the behaviour of the void to \emph{guarantee\/} certain
completeness and closure assumptions. Any behaviour of the void that
violates these assumptions will be disastrous  and render the machine
useless.

\medskip

Take, for instance, the common assumptions about time, explicit (and
sometimes) implicit in some proposed hypercomputers. Open descriptions
do not have a natural distinction between past, present and future,
since they lack regions we can easily isolate. Similarly, incomplete
descriptions naturally create undecidable statements, again making it
hard to create definite statements about time from the machine's
perspective.

If the universe forbids any openness or incompleteness regarding time,
then we could use temporal assumptions as the foundation for
completeness or closure assumptions. But if the universe does not
\emph{forbid\/} openness or incompleteness, then \emph{in general\/}
temporal inferences to bolster completeness and closure assumptions will
be invalid. We might be able to side-step these issues, for example by
using a Newtonian temporal model, but we are unlikely ever to be able to
implement such a machine in \emph{this\/} universe.

\section{Conclusion}

In summary, metamathematical arguments allow us to propose and
investigate a much wider range of formal theories than the usual
axiomatic formal theories. These more general theories, though, require
significant changes to our common assumptions regarding formal theories.

We can abandon the completeness or closure assumptions of axiomatic
formal theories, creating a theory partially determined by the `axioms'
of the system. From these theories we can also create a weaker class of
computation, which we call \emph{hypocomputation}.

Moving in the other direction is also possible, but this again requires
creating metatheories with properties not fully shared with axiomatic
formal theories. Moreover, while these hypercomputational machines may
exist in theory, implementing them is likely to prove extremely
difficult since these machines place strong requirements on the
undetermined (and undeterminable) actions of the machine.

\medskip

The behaviour of both  hypocomputational and hypercomputational machines
may seem strange in comparison to the usual computational machines.
Nonetheless, our long experience with axiomatic formal theories has
taught us a great deal about the properties of these machines. Moreover,
our experience with metamathematics has taught us how few of our
`commonsense' assumptions about the foundations of mathematics and
number theory hold as \emph{general\/} assumptions. If we are prepared
to abandon our search for universal theories (and machines), we can
readily build on these wider structures of mathematics. Nor is the act
of abandonment in any way extreme. We can (and do) build hypocomputers.
Perhaps we may yet be able to use this experience, and our knowledge of
the wider structures of mathematics, to build a true computer.

In the distant future, we may also learn how to erect machines on the
foundations of hypercomputational theory. Our inability to do so at
present, though, in no way detracts from our ability to explore the
realm of hypercomputation.


\nocite{Herken:Universal} \nocite{Stannett:Computation}

\bibliography{hypocomputation}

\begin{thebibliography}{10}

\bibitem{Ambler:Process}
Scott~W. Ambler.
\newblock (1998).
\newblock {\em Process Patterns: Building Large-Scale Systems Using Object
  Technology}.
\newblock Managing Object Technology. Cambridge University Press.

\bibitem{Atallah:Algorithms}
Mikhail~J. Atallah, editor.
\newblock (1999).
\newblock {\em Algorithms and Theory of Computation Handbook;}.
\newblock {CRC} Press.

\bibitem{Avison:Information}
David Avison and Guy Fitzgerald.
\newblock (2003).
\newblock {\em Information System Development: Methodologies, Techniques and
  Tools}.
\newblock McGraw Hill, third edition.

\bibitem{Benacerraf:Philosophy}
Paul Benacerraf and Hilary Putnam, editors.
\newblock (1964).
\newblock {\em Philosophy of Mathematics --- Selected Readings}.
\newblock Prentice-Hall Inc.

\bibitem{Berkeley:Treatise}
George Berkeley.
\newblock (1970).
\newblock {\em Treatise Concerning the Principles of Human Knowledge}.
\newblock The Bob-Merrill Company Inc.

\bibitem{Carnap:Logicist}
Rudolf Carnap.
\newblock (1964).
\newblock {\em The Logicist Foundations of Mathematics}, pages 31--41.
\newblock In Benacerraf and Putnam \cite{Benacerraf:Philosophy}.

\bibitem{Church:Unsolveable}
Alonzo Church.
\newblock (1936).
\newblock An unsolveable problem of elementary number theory.
\newblock {\em American Journal of Mathematics}, 58:345--363.

\bibitem{Clark:Paradoxes}
Michael Clark.
\newblock (2002).
\newblock {\em Paradoxes from {A} to {Z}}.
\newblock Routledge.

\bibitem{Copeland:Church}
Jack~B. Copeland.
\newblock (2000).
\newblock {\em The Church-Turing Thesis}.
\newblock Stanford University Press.

\bibitem{Curry:Remarks}
Haskell~B. Curry.
\newblock (1964).
\newblock {\em Remarks on the Definition and Nature of Mathematics}, pages
  152--156.
\newblock In Benacerraf and Putnam \cite{Benacerraf:Philosophy}.

\bibitem{Detlefsen:Hilberts}
Michael Detlefsen.
\newblock (1986).
\newblock {\em Hilbert's Program: An Essay on Mathematical Instrumentalism},
  volume 182 of {\em Studies in Epistemology}.
\newblock D. Reidel Publishing Company.

\bibitem{Gandy:Confluence}
Robin Gandy.
\newblock (1988).
\newblock {\em The Confluence of Ideas in 1936}, pages 55--111.
\newblock In Herken \cite{Herken:Universal}.

\bibitem{Glass:Defence}
Robert~L. Glass.
\newblock (September 1993).
\newblock In defence of adhocary.
\newblock {\em Journal of Systems and Software}, 22(3):149--150.

\bibitem{Herken:Universal}
Rolf Herken, editor.
\newblock (1988).
\newblock {\em The Universal Turing Machine --- {A} Half Century Survey}.
\newblock Oxford Science Publications.

\bibitem{Heyting:Intuitionist}
Arend Heyting.
\newblock (1964).
\newblock {\em The Intuitionist Foundations of Mathematics}, pages 42--49.
\newblock In Benacerraf and Putnam \cite{Benacerraf:Philosophy}.

\bibitem{Hopcroft:Formal}
John~E. Hopcroft and Jeffrey~D. Ullman.
\newblock (1969).
\newblock {\em Formal Languages and Their Relation to Automata}.
\newblock {Addison}-{Wesley} Series in Computer Science and Information
  Processing. Addison-Wesley.

\bibitem{King:Software}
David King.
\newblock (2004).
\newblock {\em Parting Software and Program Design}.
\newblock PhD thesis, University of York.

\bibitem{Kleene:Mathematical}
Stephen~Cole Kleene.
\newblock (1967).
\newblock {\em Mathematical Logic}.
\newblock John Wiley {\&} Sons Inc.

\bibitem{Kleene:Introduction}
Stephen~Cole Kleene.
\newblock (1971).
\newblock {\em Introduction to Metamathematics}, volume~1 of {\em Biblioteca
  Mathematica --- A Series, of Monographs on Pure and Applied Mathematics}.
\newblock Wolters-Noordhoff Publishing, 6\bibsuperscript{th} edition.

\bibitem{Love:Computation}
David Love.
\newblock (2006).
\newblock A note on computation.
\newblock Awaiting journal acceptance.

\bibitem{Myers:Reliable}
Glenford~J. Myers.
\newblock (1975).
\newblock {\em Reliable Software Through Composite Design}.
\newblock Mason/Charter Publishers Inc.

\bibitem{Page-Jones:Practical}
Meilir Page-Jones.
\newblock (1980).
\newblock {\em The Practical Guide to Structured Systems Design}.
\newblock Yourdon Press.

\bibitem{Podnieks:Around}
Karlis Podnieks.
\newblock (1992).
\newblock {\em Around {G}\"odel's Theorem}.
\newblock Institute of Mathematics and Computer Science, University of Latvia.

\bibitem{Sinaceur:Tarski}
Hourya Sinaceur.
\newblock (January 2001).
\newblock Alfred {T}arski: Semantic shift, heuristic shift in metamathematics.
\newblock {\em Synthese}, 126(1--2):49--65.

\bibitem{Sommerville:Engineering}
Ian Sommerville.
\newblock (1995).
\newblock {\em Software Engineering}.
\newblock Addison-Wesley, 5\bibsuperscript{th} edition.

\bibitem{Stannett:Computation}
Mike Stannett.
\newblock (2003).
\newblock Computation and hypercomputation.
\newblock {\em Minds and Machines}, 13(1):115--153.

\bibitem{Starke:Abstract}
P.~H. Starke.
\newblock (1972).
\newblock {\em Abstract Automata --- Theories of Deterministic,
  Non-Deterministic and Stochastic Automata}.
\newblock North-Holland Publishing Company.

\bibitem{Stevens:Structured}
W.~Stevens, G.~Myers, and L.~Constantine.
\newblock (1974).
\newblock Structured design.
\newblock {\em IBM Systems Journal}, 13(2):115--139.

\bibitem{Trakhtenbrot:Finite}
Borris~A. Trakhtenbrot and {Ya}.~M. Barzdin.
\newblock (1973).
\newblock {\em Finite Automata --- Behaviour and Synthesis}, volume~1 of {\em
  Fundamental Studies in Computer Science}.
\newblock North-Holland Publishing Company.

\bibitem{Turing:Computable}
Alan~M. Turing.
\newblock (November 1936).
\newblock On computable numbers, with an application to the
  \emph{Entscheidungsproblem}.
\newblock {\em Proceedings of the London Mathematical Society}, 42:230--265.
\newblock Series 2.

\bibitem{Turing:Systems}
Alan~M. Turing.
\newblock (June 1939).
\newblock Systems of logic based on ordinals.
\newblock {\em Proceedings of the London Mathematical Society},
  45(2239):161--228.

\bibitem{Heijenoort:Frege}
Jean van Heijenoort, editor.
\newblock (1967).
\newblock {\em From {Frege} to {G\"odel} --- {A} Source Book in Mathematical
  Logic}.
\newblock Harvard University Press.

\bibitem{Neumann:Axiomatization}
John von Neumann.
\newblock (1967).
\newblock {\em An Axiomatization of Set Theory}, chapter (1925), pages
  393--413.
\newblock In van Heijenoort \cite{Heijenoort:Frege}.

\bibitem{Neumann:Introduction}
John von Neumann.
\newblock (1967).
\newblock {\em On the Introduction of Transfinite Numbers}, chapter (1923),
  pages 346--354.
\newblock In van Heijenoort \cite{Heijenoort:Frege}.

\end{thebibliography}

\end{document}